\documentclass[conference]{IEEEtran}

\usepackage[numbers,square,comma,sort&compress]{natbib}
\usepackage{graphicx,amsfonts,amsmath,amssymb,array,url}
\usepackage{subfigure,setspace,multirow}
\usepackage{epstopdf}
\def\proof{\noindent\hspace{2em}{\itshape Proof: }}

\newtheorem{theorem}{Theorem}

\newcommand*{\Scale}[2][4]{\scalebox{#1}{$#2$}}%

\title{\huge Power Allocation and Cooperative Diversity in Two-Way Non-Regenerative Cognitive Radio Networks }
\author{Saeed Vahidian$^{1}$,
              Maryam Najafi$^{2}$,
              Marzieh Najafi$^{3}$,
              Fawaz S.~Al-Qahtani$^{4}$ \\
\IEEEauthorblockA{ $^{1}$ Electrical \& Computer Engineering, University of Illinois at Chicago, Chicago, USA} 
$^{2}$Electrical \& Computer Engineering, Shiraz University of Technology, Shiraz, Iran\\
$^{3}$Electrical \& Computer Engineering, K. N. Toosi University of Technology, Tehran, Iran\\
$^{4}$Electrical \& Computer Engineering Program, Texas A\textrm\&M University at Qatar, Doha, Qatar
}

\begin{document}
\maketitle
\begin{abstract}
In this paper, we investigate the performance of a dual-hop block fading cognitive radio network with underlay spectrum sharing over independent but not necessarily identically distributed (i.n.i.d.) Nakagami-$m$ fading channels. The primary network consists of a source and a destination. Depending on whether the secondary network which consists of two source nodes have a single relay for cooperation or multiple relays thereby employs opportunistic relay selection for cooperation and whether the two source nodes suffer from the primary users' (PU) interference, two cases are considered in this paper, which are referred to as Scenario (a) and Scenario (b), respectively. For the considered underlay spectrum sharing, the transmit power constraint of the proposed system is adjusted by interference limit on the primary network and the interference imposed by primary user (PU). The developed new analysis obtains new analytical results for the outage capacity (OC) and average symbol error probability (ASEP). In particular, for Scenario (a), tight lower bounds on the OC and ASEP of the secondary network are derived in closed-form. In addition, a closed from expression for the end-to-end OC of Scenario (a) is achieved. With regards to Scenario (b), a tight lower bound on the OC of the secondary network is derived in closed-form. All analytical results are corroborated using Monte Carlo simulation method.
\end{abstract}
\vspace{-1mm}
\vspace{-2mm}

\section{Introduction}
In the last decade, a significant growth in wireless cellular network has been observed~\cite{DBLP:conf/infocom/KiskaniS16, VahidMarkTCOM2016}. However, the available technologies and the limited  radio spectrum are unable to address the stringent demand for high data rate wireless communication~\cite{DBLP:journals/wicomm/KiskaniWSOG15, saeedhaj1,Vahid}. It is desired to have a solution that can mitigate the problem of inadequate cellular spectrum~\cite{sani11}. To cope with this, cognitive radio, has been proposed to alleviate the problem of resource congestion via utilization of frequency bands through dynamic spectrum management~\cite{DBLP:conf/icseng/KiskaniK11, DBLP:conf/mswim/KiskaniKV10}. There are three main cognitive radio protocols: underlay, overlay and interweave. The focus of this paper is on the underlay systems. The underlay paradigm allows secondary user (SU) utilize the spectrum as long as its interference to the primary network remains below a tolerable level~\cite{Goldsmith:2009}. 

On the other hand, cooperative communication has appeared as a powerful spatial diversity technology for wireless communications to overcome the damaging effects of multipath fading~\cite{VahidRabieiBeaulieuCommLetter}. Relay technology is one of these methods that are used to enhance the performance of wireless networks. Among various relaying schemes, best-relay selection scheme has drawn increasing research interests while being spectrally more efficient than repetition-based schemes~\cite{Bletesas:JSAC:2006}. Best relay selection~\cite{ Vahidian:WCL:2015} is an ideal protocol to reach the best performance. There are two sorts of classical relay communication protocols, i.e., non-regenerative and regenerative protocols~\cite{Vahidian:2015:WPC}.

Two-way relay networks have aroused great interest due to their potential of significantly enhancing the network throughput as well as the spectrally efficient benefits. Thereby two end users exchange information simultaneously with each other in just two time phases via a half-duplex or multiple relays~\cite{Vahidian1:TVT:2016}.

\vspace{-3mm}
\subsection{Literature on Cognitive Relaying Network}

Opportunistic relaying has been extensively studied for one-way relaying systems. For instance, in~\cite{ Duong:2012:TVT,  Shao:CommunL:2104, VahidMarkWirelessLetter}, the main purpose was to explore the effect of interference in unidirectional cognitive relay networks in terms of outage performance. In~\cite{Duong:2012:TVT}, the outage capacity (OC) of a dual-hop cognitive relay network has been explored where the power constraint of the secondary network has been derived with two different strategies. However, in this work the power constraint of the SUs depends on the knowledge of the instantaneous channel state information (CSI) of the link between the SUs and the primary users (PUs). The OC for incremental and DF relaying in underlay spectrum sharing systems over Nakagami-$m$ fading channels has been discussed in~\cite{Shao:CommunL:2104}.

Some studies have extended the opportunistic relaying to two-way relaying systems~\cite{ li:glob:2012, tao:TSP:2013, Vahidian:TVT:DF:2016, Kong:Conf:2012, pan:wcnc:2013, Saeed.Vahidian:IET:2014, saeed2}. In~\cite{ li:glob:2012} and~\cite{tao:TSP:2013}, analog network coding (ANC) and physical-layer network coding (PNC) were adopted in a two-way spectrum sharing relaying network, respectively. In~\cite{Vahidian:TVT:DF:2016}, a two-way cognitive radio system with a regenerative scheme has been studied in terms of average error rate, average sum-rate. The authors in~\cite{pan:wcnc:2013} investigated an underlay cognitive two-way relaying network in multiple access broadcasting (MABC) and  time division broadcasting (TDBC)\footnote{Two-way relaying systems can be classified into three categories: $1)$ The first model is conventional two one-way relaying to achieve a bidirectional transmission, which requires four time slots. $2)$ The alternative model is time division broadcasting (TDBC) whose number of time slots is reduced to three by using physical-layer network coding (PNC). $3)$ The final model is multiple access broadcasting (MABC) which just needs two time slots~\cite{Shahbazpanahi:TSP:2010}.}, respectively, where in~\cite{ Kong:Conf:2012} the interference caused by the PU was neglected.

Considerable efforts have been spent on addressing the performance of the cognitive radio networks. However, most of the prior works studied the performance of the cognitive radio networks for one-way relaying systems in Nakagami-$m$ environment or assumed Rayleigh fading channels. This may not be useful in a wide range of fading scenarios that are typical in realistic wireless relay applications. The performance of two-way relaying in spectrum sharing systems over Nakagami-$m$ fading channels is still much less understood. Moreover, in most of the previous works the power limitation of the secondary network depends on the knowledge of instantaneous CSI of the links between nodes~\cite{sani123, sani1234} while in some practical communication setups with high terminal speed the channel varies rapidly and experiences fast fading. Owing to these facts, the aim of this paper is to investigate the performance of dual-hop cognitive radio networks with non-regenerative relay/relays and subject to i.n.i.d. Nakagami-$m$ fading channels for two Scenarios. To our best knowledge, no similar studies have been reported despite the importance of such an issue. A detailed description of the considered Scenarios follows.

\vspace{-3mm}
\subsection{Considered Scenarios and Technical Contributions}
Scenario (a): Scenario (a) is composed of a pair of cognitive sources communicating with each other with the aid of a relay, while sharing the spectrum with a one-way primary network. In cognitive radio network, the source nodes and the relay are affected by all existed interferers and additive white Gaussian noise (AWGN). 
Scenario (b): Scenario (b) is the extension of Scenario (a) to a cognitive two-way multi-non-regenerative-relaying $({{{R}}_{{k}}}~~k = 1,...,K)$ dual-hop configuration with best relay selection strategy, where the source nodes only suffer from AWGN, while the relays are affected by all existed interferers as well as AWGN. In this Scenario the interference from the PU to the secondary sources is assumed to be neglected as in~\cite{Lee:TEC:2011}. This can be justified if the primary transmitter is located far away from the secondary sources, or the interference is modeled as the noise term~\cite{Lee:TEC:2011}.

\begin{itemize}
\item In an attempt to assess the performance, the cumulative distribution function (cdf) of the SINR at primary destination is obtained from which a closed-form expressions for the OC is derived over Nakagami-$m$ fading channels, which build the relationship between the outage performance and the related system parameters. From a realistic point of view, the choice of Nakagami-$m$ fading includes some other fading channels that are more or less severe than Rayleigh fading via the $m$ fading parameter. For example, it includes the Rayleigh fading $(m=1)$ as a special case. The Nakagami-$m$ fading also approximates the Hoyt fading, for $m<1$, and the Rician fading, for $m>1$. On the basis of the outage performance of the primary network the power allocation of the SUs are presented.

\item For both Scenarios, the exact equivalent SINR at the secondary sources are provided and upper bounded. Then, the cdfs of the upper bounded SINRs are explored. In addition, the selection of the best relay, $k$, is discussed. According to these results, tight lower bounds on the OC of Scenarios (a) and (b) are achieved. Furthermore, a closed-form expression for the end-to-end OC of the network in Scenario (a) is obtained.

\item  The tight lower bound closed-form expression for the ASEP of Scenario (a) is derived over Nakagami-$m$ fading channels.
\end{itemize}

The rest of this paper is organized as: ‎Section II, introduces channel statistics and system models. The secondary network and its performance is examined in Section III‎. Numerical results are provided in Section IV‎ and conclusions are given in Section V.
\vspace{-4mm}
\section{Channel Statistics and System Model}
\label{sec:System_Model}
Let us consider a spectrum-sharing cooperative non-regenerative relaying model. In this model, $PT$, $PX$, ${S}_i~i=1, 2$ and $R$/$R_k$ represent a primary transmitter, a primary receiver, secondary sources and secondary relay/relays, respectively. Let us consider that there is no direct link between two sources due to poor channel condition~\cite{hosseini2016real}. Further, we assume that the channels are reciprocal. A time-division multiple-access scheme is employed for the secondary nodes which is performed into two consecutive time intervals, namely the multiple-access (MA) phase and the broadcast (BC) phase. In the first phase, $PT$ sends its signal ${x_1}$, ${\rm{\mathbb{E}\{ | }}{{{x}}_1}{{{|}}^2}{{\} }} = 1$, drawn with equal probability from MPSK constellation of size $M$ to $PX$ with power ${P_{{PT}}}$. Meanwhile $S_i$~ ${i=1, 2}$ sends its information $\hat x_i$ ${\rm{\mathbb{E}\{ | }}{{{\hat x}}_i}{{{|}}^2}{{\} }} = 1,~i=1, 2$ meant for the other with transmit power of ${P_{{{S_i}}}}$ under a transmit power constraint and the relay/relays receive(s) the signals. As was previously mentioned, we assume that the instantaneous channel gain of each link ($\left| h \right|$) follows the Nakagami-$m$ distribution, where integer $m$ stands for the fading severity parameter. As such, the distribution of the ${\left| h \right|^2}$ is Gamma random variable (RV), where $\frac{{\bar \gamma }}{m}$ is scale parameter, while ${{\bar \gamma }}$ is the mean value of Gamma RVs. Therefore, we have ${\left| {{h_{PT - R}}} \right|^2}\mathop  \sim \limits^d {\rm{G}}\left( {{m_y},\frac{{{{\bar \gamma }_y}}}{{{m_y}}}} \right)$, ${\left| {{h_{R - {S_1}}}} \right|^2}\mathop  \sim \limits^d {\rm{G}}\left( {{m_w},\frac{{{{\bar \gamma }_w}}}{{{m_w}}}} \right)$, ${\left| {{h_{R - {S_2}}}} \right|^2}\mathop  \sim \limits^d {\rm{G}}\left( {{m_x},\frac{{{{\bar \gamma }_x}}}{{{m_x}}}} \right)$, ${\left| {{h_{PT - {S_1}}}} \right|^2}\mathop  \sim \limits^d {\rm{G}}\left( {{m_z},\frac{{{{\bar \gamma }_z}}}{{{m_z}}}} \right)$, and ${\left| {{h_{PT - PX}}} \right|^2}\mathop  \sim \limits^d {\rm{G}}\left( {{m_e},\frac{{{{\bar \gamma }_e}}}{{{m_e}}}} \right)$, where the symbol $\mathop  \sim \limits^d $ denotes "distributed as". Then, the signal received by ${R}$/${R_k}$ in Scenarios (a) and (b) and also by $PX$ in the MA phase can be written as
\begin{align}‎\label{gamma_SR}
\Scale[0.97]{{y_R} \hspace{-1mm}=\hspace{-1mm} \sqrt {{P_{{PT}}}} {h_{{PT} - R}}{x_1} \hspace{-1mm}+\hspace{-1mm} \sqrt {{P_{{S_1}}}} {h_{{S_1} - R}}{{\hat x}_1} \hspace{-1mm}+\hspace{-1mm} \sqrt {{P_{{S_2}}}} {h_{{S_2} - R}}{{\hat x}_2} \hspace{-1mm}+\hspace{-1mm} {n_2}},
\end{align}
\vspace{-5mm}
‎\begin{align}‎\label{Y_SR}
\Scale[0.97]{{y_{{R_k}}} \hspace{-1.2mm}=\hspace{-1.2mm} \sqrt {{P_{{PT}}}} {h_{{PT}  \hspace{-0.5mm}- \hspace{-0.5mm}{R_k}}}{x_1} \hspace{-1.2mm}+\hspace{-1.2mm} \sqrt {{P_{{S_1}}}} {h_{{S_1}  \hspace{-0.6mm}- \hspace{-0.6mm} {R_k}}}{{\hat x}_1} \hspace{-1.2mm}+\hspace{-1.2mm} \sqrt {{P_{{S_2}}}} {h_{{S_2}  \hspace{-0.6mm}-  \hspace{-0.6mm}{R_k}}}{{\hat x}_2} \hspace{-1.2mm}+ \hspace{-1.2mm}{n_{{2_k}}}},
\end{align}
\vspace{-5mm}
\begin{align}‎
\Scale[0.97]{{y_{{PX}}} \hspace{-1.2mm}=\hspace{-1.3mm} \sqrt {{P_{PT}}} {h_{{PT}  \hspace{-0.9mm}- \hspace{-0.9mm} {PX}}}{x_1}\hspace{-1.3mm} + \hspace{-1.2mm}\sqrt {{P_{{S_1}}}} {h_{{S_1}  \hspace{-0.9mm}- \hspace{-0.9mm} {PX}}}{{\hat x}_1} \hspace{-1.2mm}+ \hspace{-1.2mm}\sqrt {{P_{{S_2}}}} {h_{{S_2}  \hspace{-0.9mm}- \hspace{-0.9mm} {PX}}}{{\hat x}_2} \hspace{-1.2mm}+ \hspace{-1.2mm}{n_1}},
\label{gamma_PD1}
\end{align}‎‎‎where ${h_{{{PT - PX}}}}$‎, ‎${h_{{{S_1 - PX}}}}$‎, ‎${h_{{{S_2 - PX}}}}$‎, ‎${h_{PT - R/{R_k}}}$‎, ‎${h_{S_1- R/{R_k}}}$ and ${h_{S_2 - R/{R_k}}}$ are the channel coefficients of ${{PT}} \hspace{-1.3mm} \to\hspace{-1.3mm} {{PX}},{\mkern 1mu} {\mkern 1mu} \,{{S_1}} \hspace{-1.3mm}\to\hspace{-1.3mm} {{PX}},{\mkern 1mu} {\mkern 1mu} \,{{S_2}} \hspace{-1.3mm}\to \hspace{-1.3mm} {{PX}},{\mkern 1mu} \,PT \to R/{R_k}‎$, $‎{S_1} \to R/{R_k}$‎ and ‎${S_2} \to R/{R_k}$ links, respectively, while ${n_1}$ and ${n_2}/{n_{{2_k}}}$ are AWGN at the primary receiver terminal and the relay/relays, respectively‎. Noise signals at all nodes are zero mean with power spectral density of ${N_0}$.
In the second phase, PT, sends another information message $x_2$~${\rm{\mathbb{E}\{ | }}{{{x}}_2}{{{|}}^2}{{\} }} = 1$ to $PX$. Meanwhile in the second phase, $\rm{R}$/${{\rm{R}}_k}$, simply scales the received signal before broadcasting to the destination with transmit power ${P_R}/{P_{{R_k}}}$. Accordingly, the received signal at $PX$ and $S_1$ in Scenario (a) and (b) are given respectively by
\begin{align}‎
&\Scale[0.97]{{y_{{PX}}} = \sqrt {{P_{{\rm{PT}}}}} {h_{{\rm{PT - PX}}}}{x_2}+ {{{G}}_k}\sqrt {{P_{{\rm{}}{{\rm{R}}_k}}}} {h_{{\rm{}}{{\rm{R}}_k}{\rm{ - PD}}}}{y_{{{\rm{R}}_k}}} + {n_{\rm{1}}}},
\label{Y_PX2}
\end{align}
\vspace{-5mm}
\begin{align}‎
&\Scale[0.97]{{y_{{S_1}}} = \sqrt {{P_{PT}}} {h_{PT - {S_1}}}{x_2} + G\sqrt {{P_R}} \sqrt {{P_{PT}}} {h_{PT - R}}{h_{R - {S_1}}}{x_1}}
 \notag \\
 &\;+ \Scale[0.97]{G\sqrt {{P_R}} \sqrt {{P_{{S_2}}}} {h_{{S_2} - R}}{h_{R - {S_1}}}{{\hat x}_2} + G\sqrt {{P_R}} {h_{R - {S_1}}}{n_2} + {n_3}},\label{Y_ST1}
\end{align}
\vspace{-5mm}
\begin{align}‎
&\Scale[0.97]{{y_{{S_{1,\,k}}}} = {G_k}\sqrt {{P_{{R_k}}}} \sqrt {{P_{PT}}} {h_{{R_k} - {S_1}}}{h_{PT - {R_k}}}{x_1} +
{G_k}\sqrt {{P_{{R_k}}}}} \notag
\\
&\Scale[0.97]{\times\sqrt {{P_{{S_2}}}} {h_{{R_k} - {S_1}}}{h_{{S_2} - {R_k}}}{{\hat x}_2} \hspace{-1mm}+\hspace{-1mm} {G_k}\sqrt {{P_{{R_k}}}} {h_{{R_k} - {S_1}}}{n_{{2_k}}} \hspace{-1.2mm}+\hspace{-1.2mm} {n_{3_k}}},
\label{Y_ST1_Selection}
\end{align}‎where ‎${h_{{{PT - S_1}}}}$ is the channel coefficient of ${\mkern 1mu} {{PT}} \hspace{-1.3mm}\to\hspace{-1.3mm} {{S_1}}$ link, while ${n_3}$‎/${n_{3_k}}$, is AWGN at $S_1$/$S_{1,k}$‎ and $G$ and $G_k$, are the relay amplification gains defined as
\begin{align}‎
&\Scale[0.93]{{G^{ - 1}}\hspace{-0.5mm} \buildrel \Delta \over =\hspace{-0.5mm} \sqrt {{P_{PT}}{{\left| {{h_{PT - R}}} \right|}^2} \hspace{-0.5mm}+ \hspace{-0.5mm} {P_{{S_1}}}{{\left| {{h_{{S_1} - R}}} \right|}^2} \hspace{-0.5mm}+\hspace{-0.5mm} {P_{{S_2}}}{{\left| {{h_{R - {S_2}}}} \right|}^2} \hspace{-0.5mm}+\hspace{-0.5mm} {N_0}}}, \label{G}
\end{align}
\vspace{-1mm}
\begin{align}‎
\Scale[0.93]{G_k^{ - 1} \hspace{-1mm}\buildrel \Delta \over = \hspace{-1mm}{\rm{ }}\sqrt {{P_{PT}}{{\left| {{h_{PT - {R_k}}}} \right|}^2} \hspace{-1mm}+\hspace{-1mm} {P_{{S_1}}}{{\left| {{h_{{S_1} - {R_k}}}} \right|}^2} \hspace{-1mm}+\hspace{-1mm} {P_{{S_2}}}{{\left| {{h_{{R_k} - {S_2}}}} \right|}^2} \hspace{-1mm}+\hspace{-1mm} {N_0}}}. \label{G_k}
\end{align}

In the following sections, we will evaluate the network performance in key operation metrics.

‎\section{Performance Analysis}‎
In this section the statistics of the equivalent SINR will be explored, tight lower bounds on the most important performance metrics of the secondary network will be quantified and some useful observations in this realm will be provided.

\vspace{-4mm}
‎\subsection{Primary OC and Power Constraint of the Secondary Nodes}‎
In this subsection, we present some of statistics for the OC of the primary network. The OC is defined as the probability that the instantaneous SINRs falls below a predetermined rate $R$ bits/sec/Hz. Hence the primary OC can be explored by the following theorem:
\begin{theorem} \label{Thm-outage probability-DMM}
\emph{The OC of the primary network is obtained in closed-form expression as shown in \eqref{OP of PN},}

\begin{figure*}
‎\begin{align}‎
&\Scale[0.97]{{P_{{\rm{out,}}\,{\rm{pri}}}} = 1 -
\frac{{\left( {{m_e} - 1} \right)!{{\left( {{m_f}} \right)}^{{m_f}}}{{\left( {{m_g}} \right)}^{{m_g}}}}}{{\Gamma \left( {{m_e}} \right)\Gamma \left( {{m_g}} \right){{\left( {{{\bar \gamma }g}} \right)}^{{m_g}}}\Gamma \left( {{m_f}} \right){{\left( {\bar \gamma f^{}} \right)}^{{m_f}}}}} \exp \left( { - \left( {\frac{{{m_e}}}{{{{\bar \gamma }e}{{\bar \gamma }{\rm{P}}}}}{\gamma {{\rm{th}}}}} \right)} \right) \sum\limits_{\ell  = 0}^{{m_e} - 1} {\sum\limits_{{\tau 1=0}}^\ell  {\sum\limits_{{\tau 2=0}}^{\ell  - {\tau 1}} {\frac{{{{\left( {\frac{{{m_e}{\gamma {{\rm{th}}}}}}{{{{\bar \gamma }e}{{\bar \gamma }{\rm{P}}}}}} \right)}^\ell }{{\left( {{{\bar \gamma }{{{\rm{S}}_1}}}} \right)}^{{\tau 1}}}{{\left( {{{\bar \gamma }{{{\rm{S}}_2}}}} \right)}^{{\tau _2}}}}}{{\ell !}}} } }}
\nonumber\\&
\Scale[0.97]{\times \left( \begin{array}{l}
\ell  - {\tau _1}\\
\,\,{\tau _2}
\end{array} \right)\,\left( \begin{array}{l}
\ell \\
{\tau _1}
\end{array} \right)  \Gamma \left( {{m_f} + {\tau _1}} \right)
\Gamma \left( {{m_g} + {\tau 2}} \right){\left( {\frac{{{m_f}}}{{{{\bar \gamma }f}}} + \frac{{{m_e}{{\bar \gamma }{{{\rm{S}}_1}}}}}{{{{\bar \gamma }e}{{\bar \gamma }{\rm{P}}}}}{\gamma {{\rm{th}}}}} \right)^{ - \left( {{m_f} + {\tau 1}} \right)}}{\left( {\frac{{{m_g}}}{{{{\bar \gamma }g}}} + \frac{{{m_e}{{\bar \gamma }{{{\rm{S}}_2}}}}}{{{{\bar \gamma }e}{{\bar \gamma }{\rm{P}}}}}{\gamma {{\rm{th}}}}} \right)^{ - \left( {{m_g} + {\tau _2}} \right)}}},
\label{OP of PN}
\end{align}
\rule{\linewidth}{1pt}
\vspace{-7mm}
\end{figure*}

\noindent where ${\gamma _{{\rm{th}}}}=2^{R_{\rm{P}}}-1$ and ${R_{\rm{P}}}$ is the primary transmission rate.

\end{theorem}
\proof
\emph{See Appendix~\ref{apx1}}\\
In both Scenarios, by assuming ${\bar \gamma _{{S_1}}} = {\bar \gamma _{{S_2}}}$ and without loss of generality, the power limitation of $S_i,~i=1, 2$ are derived under the constraint which guarantees that the interference on $PX$ does not exceed a threshold. Therefore, the power constraint of $S_i~i=1, 2$ is obtained ‎by solving ${P_{{\rm{out,}}\,{\rm{pri}}}} \le P_{{\rm{out}},{\rm{pri}}}^{{\rm{Thr}}}$ w.r.t. ${P _{{S_1}}} $.

Similar to \emph{Theorem~\ref{Thm-outage probability-DMM}} and according to \eqref{Y_PX2} the transmit power of $k$th secondary relay regulated alone by the primary outage constraint can be readily achieved by solving \eqref{OP of PN2} w.r.t. ${P _{{R}}} $.
‎\begin{align}‎
&\Scale[0.97]{{P_{{\rm{out,}}\,{\rm{{pri}_2}}}} =
1 - \Gamma(m_e)\exp \left( { - \frac{{{m_e}\Theta }}{{{{\bar \gamma }_e}{{\bar \gamma }_{\rm{P}}}}}} \right){\sum\limits_{\ell  = 0}^{{m_e} - 1} {\left( { - \frac{{{m_e}\Theta }}{{{{\bar \gamma }_e}{{\bar \gamma }_{\rm{P}}}}}} \right)} ^\ell }\frac{{{{\left( {{m_l}} \right)}^{{m_l}}}}}{{\Gamma \left( {{m_l}} \right)\bar \gamma _l^{{m_l}}}} }\nonumber\\&
\Scale[0.97]{\frac{{\sum\limits_{i = 0}^\ell {l \choose i} }}{{\ell !}}\left( {{m_l} - 1 + \ell  - i} \right)!{\left( {\frac{{{m_e}{{\bar \gamma }_{\rm{R}}}\Theta }}{{{{\bar \gamma }_e}{{\bar \gamma }_{\rm{P}}}}} + \frac{{{m_l}}}{{{{\bar \gamma }_l}}}} \right)^{ - \left( {{m_l} + \ell  - i} \right)}}  \le P_{{\rm{out}},{\rm{pri}}}^{{\rm{Thr}}}}.
\label{OP of PN2}
\end{align}
\vspace{-10mm}

‎\subsection{cdf of the Equivalent SINRs}
By substituting \eqref{G} into \eqref{Y_ST1} and after some algebraic manipulations of the SINR at $S_1$ in Scenario (a) can be obtained as in Eq. (11) of ~\cite{Vahidian1:TVT:2016}. To simplify the ensuing derivations, ${\gamma _{{{S_1}}}}$ and ${\gamma _{{S_1},\,k}}$ should be addressed in more mathematically tractable forms. As such, we use the tight upper bounds for ${\gamma _{{{S_1}}}}$ and ${\gamma _{{S_1},\,k}}$ as presented in Eqs. (9) and (24) of~\cite{Vahidian1:TVT:2016}. In the next step we proceed to explore the OC of both Scenarios.

\subsubsection{cdf of the SINR of Scenario (a)}
The closed-form of the cdf of $\gamma _{{{S_1}}}^{{\rm{up}}}{\mkern 1mu}$  for Scenario (a) can be presented in the following key result.
\begin{theorem}\label{op-secanrio-a}
\emph{The cdf of the SINR of Scenario (a) is obtained in closed-form expression as shown in \eqref{CDF}}
\begin{figure*}
\vspace{-2mm}
‎\begin{align}‎
&\Scale[0.97]{{F_{\gamma _{{S_1}}^{{\rm{up}}}}}\left( \Theta  \right)\approx 1 - \left( {{m_x} - 1} \right)!\left( {{m_w} - 1} \right)!\frac{{{{\left( {{m_Y}} \right)}^{{m_Y}}}{{\left( {{m_z}} \right)}^{{m_z}}}{{\left( {{m_z}} \right)}^{{m_z}}}}}{{\Gamma \left( {{m_x}} \right)\Gamma \left( {{m_y}} \right)\Gamma \left( {{m_z}} \right)\Gamma \left( {{m_w}} \right)\Gamma \left( {{m_z}} \right){{\left( {\frac{{{{\bar \gamma }{\rm{R}}}{{\bar \gamma }{\rm{P}}}}}{{{{\bar \gamma }{\rm{S}}}}}{{\bar \gamma }y}} \right)}^{{m_Y}}}{{\left( {{{\bar \gamma }{\rm{P}}}{{\bar \gamma }_z}} \right)}^{2{m_z}}}}} }\nonumber\\&
\Scale[0.97]{\times \exp \left( { - \left( {\frac{{\left( {\frac{{{{\bar \gamma }{\rm{R}}}}}{{{{\bar \gamma }{\rm{S}}}}} + 1} \right){m_x}\Theta }}{{{{\bar \gamma }x}{{\bar \gamma }{\rm{R}}}}} + \frac{{{m_w}\Theta }}{{{{\bar \gamma }W}{{\bar \gamma }{\rm{R}}}}}} \right)} \right) \times \sum\limits_{n = 0}^{{m_x} - 1} {\sum\limits_{{i_1} = 0}^n {\sum\limits_{{i_2} = 0}^{n - {i_1}} {\sum\limits_{k = 0}^{{m_w} - 1} {\sum\limits_{{k_1} = 0}^k {\left( \begin{array}{l}
k\\
{k_1}
\end{array} \right)} } \left( \begin{array}{l}
n\\
{i_1}
\end{array} \right)} } \left( \begin{array}{l}
n - {i_1}\\
\,\,\,{i_2}
\end{array} \right)}  \times} \nonumber\\&
\Scale[0.95]{\frac{{{{\left( {\frac{{{{\bar \gamma }{\rm{R}}}}}{{{{\bar \gamma }{\rm{S}}}}} + 1} \right)}^{n - {i_1} - {i_2}}}{{\left( {\frac{{{m_x}\Theta }}{{{{\bar \gamma }x}{{\bar \gamma }{\rm{R}}}}}} \right)}^n}{{\left( {\frac{{{m_w}\Theta }}{{{{\bar \gamma }W}{{\bar \gamma }{\rm{R}}}}}} \right)}^k}}}{{n!k!}}\left( {{m_y} + {i_1} - 1} \right)!\left( {{m_z} + {i_2} - 1} \right)!\left( {{m_z} + {k_1} - 1} \right)!{\left( {\frac{{{m_w}\Theta }}{{{{\bar \gamma }W}{{\bar \gamma }{\rm{R}}}}} + \frac{{{m_z}}}{{{{\bar \gamma }{\rm{P}}}{{\bar \gamma }z}}}} \right)^{ - \left( {{m_z} + {k_1}} \right)}} } \nonumber\\&
\Scale[0.97]{\times{\left( {\frac{{{m_x}\Theta }}{{{{\bar \gamma }x}{{\bar \gamma }{\rm{R}}}}} + \frac{{{m_z}}}{{{{\bar \gamma }{\rm{P}}}{{\bar \gamma }z}}}} \right)^{ - \left( {{m_z} + {i_2}} \right)}}\,{\left( {\frac{{{m_x}\Theta }}{{{{\bar \gamma }x}{{\bar \gamma }{\rm{R}}}}} + \frac{{{m_y}}}{{\frac{{{{\bar \gamma }{\rm{R}}}{{\bar \gamma }{\rm{P}}}}}{{{{\bar \gamma }{\rm{S}}}}}{{\bar \gamma }y}}}} \right)^{ - \left( {{m_y} + {i_1}} \right)}}},
\label{CDF}
‎\end{align}‎‎
\rule{\linewidth}{1pt}
\vspace{-7mm}
\end{figure*}
where $\Theta$ is the outage threshold.
\end{theorem}
\proof
\emph{See Appendix~\ref{apx2}.}

\subsubsection{End-to-End cdf of the SINR of Scenario (a)}
The end-to-end cdf of the SINR of Scenario (a) is derived as

‎\begin{align}
{F_{\gamma _{{\rm{e2e}}}^{{\rm{up}}}}}\left( \Theta  \right) \approx 1 - \left[\Upsilon _1+\Upsilon _2 \right] \left[\Upsilon _3+\Upsilon _4 \right]
‎\end{align}‎‎
where $\Upsilon _1$ and $\Upsilon _2$ are given respectively in \eqref{Upsilon1} and \eqref{Upsilon2}. The terms $\Upsilon _3$ and $\Upsilon _4$ can be obtained respectively by
replacing appropriately the parameters in $\Upsilon _1$ and $\Upsilon _2$, i.e., ${m_v} \leftrightarrow {m_z}$, ${m_x} \leftrightarrow {m_w}$, ${\bar \gamma _x} \leftrightarrow {\bar \gamma _w}$ and ${\bar \gamma _v} \leftrightarrow {\bar \gamma _z}$.
\begin{figure*}
\vspace{-2mm}
\small
‎\begin{align}\label{Upsilon1}
&\Scale[0.97]{\Upsilon 1 =\frac{{{{\left( {{m_y}} \right)}^{{m_y}}}{{\left( {{m_z}} \right)}^{{m_z}}}}}{{\Gamma \left( {{m_y}} \right)\Gamma \left( {{m_z}} \right){{\left( {{{\bar \gamma }{\rm{P}}}{{\bar \gamma }z}} \right)}^{{m_z}}}{{\left( {\frac{{{{\bar \gamma }{\rm{R}}}{{\bar \gamma }{\rm{P}}}}}{{{{\bar \gamma }{\rm{S}}}}}{{\bar \gamma }y}} \right)}^{{m_y}}}}}\rm{exp}\left( { - \left( {\frac{{{{\bar \gamma }{\rm{R}}}}}{{{{\bar \gamma }{\rm{S}}}}} + 1} \right)} \right)\sum\limits_{\rho  = 0}^{{m_x} - 1} {\sum\limits_{{\varepsilon 1} = 0}^\rho  {\sum\limits_{{\varepsilon 2} = 0}^{\rho  - {\varepsilon 1}} {\left( \begin{array}{l}
\rho \\
{\varepsilon _1}
\end{array} \right)} } \left( \begin{array}{l}
\rho  - {\varepsilon _1}\\
\,\,\,{\varepsilon _2}
\end{array} \right)} } \nonumber\\&
\Scale[0.97]{\times {{\left( {\frac{{{{\bar \gamma }{\rm{R}}}}}{{{{\bar \gamma }{\rm{S}}}}} + 1} \right)}^{\rho  - {\varepsilon 1} - {\varepsilon 2}}} \frac{{{{\left( {\frac{{{m_x}\Theta }}{{{{\bar \gamma }x}{{\bar \gamma }{\rm{R}}}}}} \right)}^\rho }}}{{\rho !}}\Gamma \left( {{m_z} + {\varepsilon 2}} \right){\left( {\frac{{{m_x}\Theta }}{{{{\bar \gamma }x}{{\bar \gamma }{\rm{R}}}}} + \frac{{{m_z}}}{{{{\bar \gamma }{\rm{P}}}{{\bar \gamma }z}}}} \right)^{{m_z} + {\varepsilon 2}}}\left( {{m_y} + {\varepsilon 1} - 1} \right)!{\left( {\frac{{{m_x}\Theta }}{{{{\bar \gamma }x}{{\bar \gamma }{\rm{R}}}}} + \frac{{{m_y}}}{{\frac{{{{\bar \gamma }{\rm{R}}}{{\bar \gamma }{\rm{P}}}}}{{{{\bar \gamma }{\rm{S}}}}}{{\bar \gamma }_y}}}} \right)^{{m_y} + \varepsilon }} - }\nonumber\\&
\Scale[0.95]{\frac{{{{\left( {{m_y}} \right)}^{{m_y}}}{{\left( {{m_z}} \right)}^{{m_z}}}}}{{\Gamma \left( {{m_y}} \right)\Gamma \left( {{m_z}} \right){{\left( {{{\bar \gamma }{\rm{P}}}{{\bar \gamma }z}} \right)}^{{m_z}}}{{\left( {\frac{{{{\bar \gamma }{\rm{R}}}{{\bar \gamma }{\rm{P}}}}}{{{{\bar \gamma }{\rm{S}}}}}{{\bar \gamma }y}} \right)}^{{m_y}}}}}\exp \left( { - \left( {\frac{{{m_x}\Theta }}{{{{\bar \gamma }x}{{\bar \gamma }{\rm{R}}}}}\left( {\frac{{{{\bar \gamma }{\rm{R}}}}}{{{{\bar \gamma }{\rm{S}}}}} + 1} \right)} \right) + \left( {\frac{{{m_v}{{\bar \gamma }{\rm{R}}}}}{{{{\bar \gamma }{\rm{P}}}{{\bar \gamma }v}{{\bar \gamma }{\rm{S}}}}}} \right)} \right) } \nonumber\\&
\Scale[0.97]{\times\sum\limits_{\varpi  = 0}^{{m_v} - 1} {\sum\limits_{\rho  = 0}^{{m_x} - 1} {\sum\limits_{{\varepsilon 1} = 0}^\rho  {\sum\limits_{{\varepsilon 2} = 0}^{\rho  - {\varepsilon _1}} {\sum\limits_{{t_1} = 0}^\varpi  {\sum\limits_{{t_2} = 0}^{\varpi  - {t_1}} {\left( \begin{array}{l}
\varpi \\
{t_1}
\end{array} \right)} } \left( \begin{array}{l}
\varpi  - {t_1}\\
\,\,\,{t_2}
\end{array} \right)\left( \begin{array}{l}
\rho \\
{\varepsilon _1}
\end{array} \right)} } \left( \begin{array}{l}
\rho  - {\varepsilon _1}\\
\,\,\,{\varepsilon _2}
\end{array} \right){{\left( {\frac{{{{\bar \gamma }{\rm{R}}}}}{{{{\bar \gamma }{\rm{S}}}}} + 1} \right)}^{\rho  - {\varepsilon 1} - {\varepsilon 2}}}{{\left( {\frac{{{{\bar \gamma }{\rm{R}}}}}{{{{\bar \gamma }{\rm{S}}}}}} \right)}^{\varpi  - {t_1} - {t_2}}} \times } }}\nonumber\\&
\Scale[0.97]{\frac{{{{\left( {\frac{{{m_v}}}{{{{\bar \gamma }{\rm{P}}}{{\bar \gamma }v}}}} \right)}^\varpi }{{\left( {\frac{{{m_x}\Theta }}{{{{\bar \gamma }x}{{\bar \gamma }{\rm{R}}}}}} \right)}^\rho }}}{{\varpi !\rho !}}\Gamma \left( {{t_2} + {\varepsilon 2} + {m_z}} \right){\left( {\frac{{{m_x}\Theta }}{{{{\bar \gamma }x}{{\bar \gamma }{\rm{R}}}}} + \frac{{{m_v}}}{{{{\bar \gamma }{\rm{P}}}{{\bar \gamma }v}}} + \frac{{{m_z}}}{{{{\bar \gamma }{\rm{P}}}{{\bar \gamma }z}}}} \right)^{-\left( {t_2} + {\varepsilon 2} + {m_z}\right)}}\Gamma \left( {{t_1} + {\varepsilon _1} + {m_y}} \right)  }
\Scale[0.97]{\times{\left( {\frac{{{m_x}\Theta }}{{{{\bar \gamma }x}{{\bar \gamma }{\rm{R}}}}} + \frac{{{m_v}}}{{{{\bar \gamma }{\rm{P}}}{{\bar \gamma }v}}} + \frac{{{m_y}}}{{\frac{{{{\bar \gamma }{\rm{R}}}{{\bar \gamma }{\rm{P}}}}}{{{{\bar \gamma }{\rm{S}}}}}{{\bar \gamma }y}}}} \right)^{-\left( {t_1} + {\varepsilon _1} + {m_y}\right)}}},
‎\end{align}‎‎
\rule{\linewidth}{1pt}
\vspace{-4mm}
\end{figure*}
\begin{figure*}
\vspace{-5mm}
\small
‎\begin{align}\label{Upsilon2}
&\Scale[0.99]{\Upsilon _2=
\frac{{{{\left( {{m_v}} \right)}^{{m_v}}}}}{{\Gamma \left( {{m_v}} \right){{\left( {{{\bar \gamma }{\rm{P}}}{{\bar \gamma }v}} \right)}^{{m_v}}}}}\frac{{{{\left( {{m_x}} \right)}^{{m_x}}}}}{{\bar \gamma x^{{m_x}}}}\exp \left( { - \frac{{{m_x}\Theta }}{{{{\bar \gamma }x}{{\bar \gamma }{\rm{R}}}}}} \right)\exp \left( { - \frac{{{{\bar \gamma }{\rm{R}}}}}{{{{\bar \gamma }{\rm{S}}}}}\left( {\frac{{{m_x}\Theta }}{{{{\bar \gamma }x}{{\bar \gamma }{\rm{R}}}}} + \frac{{{m_v}}}{{{{\bar \gamma }{\rm{P}}}{{\bar \gamma }v}}}} \right)} \right)\frac{{{{\left( {{m_y}} \right)}^{{m_y}}}}}{{\Gamma \left( {{m_y}} \right){{\left( {\frac{{{{\bar \gamma }{\rm{R}}}{{\bar \gamma }{\rm{P}}}}}{{{{\bar \gamma }{\rm{S}}}}}{{\bar \gamma }_y}} \right)}^{{m_y}}}}}} \nonumber\\&
\Scale[0.99]{\times\frac{{{{\left( {{m_z}} \right)}^{{m_z}}}}}{{\Gamma \left( {{m_z}} \right){{\left( {{{\bar \gamma }{\rm{P}}}{{\bar \gamma }z}} \right)}^{{m_z}}}}}\sum\limits_{i = 0}^{{m_x} - 1} {\sum\limits_{j = 0}^i {\sum\limits_{k = 0}^{{m_v} + j - 1} {\sum\limits_{{k_1} = 0}^k {\sum\limits_{{k_2} = 0}^{{k_1}} {\left( \begin{array}{l}
{k_1}\\
{k_2}
\end{array} \right)} \left( \begin{array}{l}
k\\
{k_1}
\end{array} \right)} \,{{\left( {\frac{{{{\bar \gamma }{\rm{R}}}}}{{{{\bar \gamma }{\rm{S}}}}}} \right)}^{k - {k_1}}}\left( \begin{array}{l}
i\\
j
\end{array} \right)} }  }  } \nonumber\\&
\Scale[0.99]{\times\frac{{\Gamma \left( {{m_v} + j} \right){{\left( {\frac{\Theta }{{{{\bar \gamma }{\rm{R}}}}}} \right)}^i}\Gamma \left( {{k_2} + {m_z}} \right)\Gamma \left( {{k_1} + {m_y} - {k_2}} \right)}}{{\left( i \right)!\left( k \right)!{{\left( {\frac{{{m_x}}}{{{{\bar \gamma }x}}}} \right)}^{{m_x} - i}}{{\left( {\frac{{{m_x}\Theta }}{{{{\bar \gamma }x}{{\bar \gamma }{\rm{R}}}}} + \frac{{{m_v}}}{{{{\bar \gamma }{\rm{P}}}{{\bar \gamma }v}}}} \right)}^{{m_v} + j - k}}}} {\left( {\frac{{{m_z}}}{{{{\bar \gamma }{\rm{P}}}{{\bar \gamma }z}}} + \frac{{{m_x}\Theta }}{{{{\bar \gamma }x}{{\bar \gamma }{\rm{R}}}}} + \frac{{{m_v}}}{{{{\bar \gamma }{\rm{P}}}{{\bar \gamma }v}}}} \right)^{ - \left( {{k_2} + {m_z}} \right)}}
{\left( {\frac{{{m_y}}}{{\frac{{{{\bar \gamma }{\rm{R}}}{{\bar \gamma }{\rm{P}}}}}{{{{\bar \gamma }{\rm{S}}}}}{{\bar \gamma }y}}} + \frac{{{m_x}\Theta }}{{{{\bar \gamma }x}{{\bar \gamma }{\rm{R}}}}} + \frac{{{m_v}}}{{{{\bar \gamma }{\rm{P}}}{{\bar \gamma }v}}}} \right)^{ - \left( {{k_1} + {m_y} - {k_2}} \right)}}.}
‎\end{align}‎
\rule{\linewidth}{1pt}
\vspace{-5mm}
\end{figure*}

\subsubsection{cdf of the SINR of Scenario (b)}
In general, by applying the best relay selection strategy, the relay which has the highest value of end-to-end SINR is viewed as the "best" one. Hence, the end-to-end received SINR at terminals is given by
‎\begin{align}
\gamma _{{\rm{e2e}}}^{{\rm{up}}} \hspace{-1mm}=\hspace{-1mm} \mathop {\max }\limits_{k = 1,2,...,K} \left\{ {\gamma _{}^{{\rm{up}}}} \right\} = \hspace{-1mm}\mathop {\max }\limits_{k = 1,2,...,K} \left\{ {\min \left( {\gamma _{{{S_{1,k}}}}^{{\rm{up}}},\gamma _{{{S_{2,k}}}}^{{\rm{up}}}} \right)} \right\}.
‎\end{align}‎
Since the exact cdf of the end-to-end SINR is intractable, in the following, we obtain the cdf of the end-to-end upper bounded SINR.
\begin{theorem}\label{op-secanrio-b}
\emph{The cdf of the end-to-end SINR of Scenario (b) is derived  in \eqref{CDF_Selection}.}

\begin{figure*}
\vspace{-2mm}
\small
‎\begin{align}‎
&\Scale[0.95]{{F_{\gamma _{{\rm{e2e}}}^{{\rm{up}}}}}\left( \Theta  \right) =
\sum\limits_{k = 0}^K {\frac{{{{\left( { - 1} \right)}^k}}}{{k!}}} \,\,\,\,\sum\limits_{{\varpi 1},...,{\varpi k}}^K {\,\,\prod\limits_{\tau  = 1}^k {\left[ {\frac{{{{\left( {{m_{{y_{{\varpi \tau }}}}}} \right)}^{{m_{{y_{{\varpi \tau }}}}}}}}}{{\Gamma \left( {{m_{{y_{{\varpi \tau }}}}}} \right){{\left( {\frac{{{{\bar \gamma }{\rm{R}}}{{\bar \gamma }{\rm{P}}}}}{{{{\bar \gamma }{\rm{S}}}}}{{\bar \gamma }{{y_{{\varpi \tau }}}}}} \right)}^{{m_{{y_{{\varpi _\tau }}}}}}}}}} \right. \times } } }\nonumber\\&
\Scale[0.97]{\exp \left( { - \left( {\left( {\frac{{{m_{{x_{{\varpi \tau }}}}}}}{{{{\bar \gamma }{{x_{{\varpi \tau }}}}}{{\bar \gamma }{\rm{R}}}}}} \right)\left( {\frac{{{{\bar \gamma }{\rm{R}}}}}{{{{\bar \gamma }{\rm{S}}}}} + 1} \right) + \left( {\frac{{{m_{{w_{{\varpi \tau }}}}}}}{{{{\bar \gamma }{{w_{{\varpi \tau }}}}}{{\bar \gamma }{\rm{R}}}}}} \right)\left( {\frac{{{{\bar \gamma }{\rm{R}}}}}{{{{\bar \gamma }{\rm{S}}}}} + 1} \right)} \right)\Theta } \right) \times \sum\limits_{n = 0}^{{m_{{x_k}}} - 1} {\sum\limits_{{n_1} = 0}^{{m_{{x_k}}} - 1} {\sum\limits_{{i_1} = 0}^{n + {n_1}} {\sum\limits_{{i_2} = 0}^{n + {n_1} - {i_1}} {\frac{{\Gamma \left( {{i_1} + {m_{{y_{{\varpi _\tau }}}}}} \right)}}{{{n_1}!n!}}} } } } }\nonumber\\&
\Scale[0.97]{ \times {{\left( {\frac{{{m_{{w_{{\varpi \tau }}}}}\Theta }}{{{{\bar \gamma }{{w_{{\varpi \tau }}}}}{{\bar \gamma }{\rm{R}}}}}} \right)}^{{n_1}}} {\left( {\frac{{{m_{{x_{{\varpi \tau }}}}}\Theta }}{{{{\bar \gamma }{{x_{{\varpi \tau }}}}}{{\bar \gamma }{\rm{R}}}}}} \right)^n}\binom{n+n_1}{i_1}\binom{n + {n_1} - {i_1}}{i_2}  {\left( {\frac{{{{\bar \gamma }{\rm{R}}}}}{{{{\bar \gamma }{\rm{S}}}}}} \right)^{{i_2}}}\left. {{{\left( {\frac{{{m_{{x_{{\varpi \tau }}}}}\Theta }}{{{{\bar \gamma }{{x_{{\varpi \tau }}}}}{{\bar \gamma }{\rm{R}}}}} + \frac{{{m_{{w_{{\varpi \tau }}}}}\Theta }}{{{{\bar \gamma }{{w_{{\varpi \tau }}}}}{{\bar \gamma }{\rm{R}}}}} + \frac{{{{\bar \gamma }{\rm{S}}}{m_{{y_{{\varpi \tau }}}}}}}{{{{\bar \gamma }{\rm{R}}}{{\bar \gamma }{\rm{P}}}{{\bar \gamma }{{y_{{\varpi \tau }}}}}}}} \right)}^{ - \left( {{i_1} + {m_{{y_{{\varpi _\tau }}}}}} \right)}}} \right]}.
\label{CDF_Selection}
‎\end{align}‎‎
\rule{\linewidth}{1pt}
\end{figure*}
\vspace{-.5mm}
\end{theorem}
\proof
\emph{See Appendix~\ref{apx3}.}

By using \eqref{CDF} and \eqref{CDF_Selection}, we can now obtain the following lower bounds on the exact OC of the Scenarios (a) and (b) respectively as
‎\begin{align}‎
{P_{{\rm{ou}}{{\rm{t}}_{{\rm{e2e}}}}}} = {F_{\gamma _{{\rm{e2e}}}^{{\rm{up}}}}}\left( {{\gamma _{{\rm{th}}}}} \right),\label{OP-Scenario a}\\
{P_{{\rm{ou}}{{\rm{t}}_{{S_1}}}}} = {F_{\gamma _{{{\rm{S}}_{\rm{1}}}}^{{\rm{up}}}}}\left( {{\gamma _{{\rm{th}}}}} \right).\label{OP-Scenario b}
‎\end{align}‎

‎\vspace{-12.5mm}‎
\subsection{Error Probability‎}


‎In this subsection we concentrate on the ASEP of the secondary network for Scenario (a) as another important performance evaluation metric. ‎For several modulation schemes, in practical systems ASEP can be lower bounded as~\cite[Eq. (9)]{Duong:Let:2013}. The following theorem summarizes the ASEP of Scenario (a).
 \begin{theorem}
\emph{We derive a closed-form expression for the ASEP of Scenario (a) as in \eqref{SEP_Final} shown on the top of the next page,}
\begin{figure*}
\vspace{-2mm}
\small
‎\begin{align}‎
&\Scale[0.98]{\bar P_{{S_1}}^e = \frac{a}{2} - \frac{{a\sqrt b }}{{2\sqrt \pi  }}\left( {{m_x} - 1} \right)!\left( {{m_w} - 1} \right)!\frac{{{{\left( {{m_Y}} \right)}^{{m_Y}}}{{\left( {{m_z}} \right)}^{{m_z}}}{{\left( {{m_z}} \right)}^{{m_z}}}}}{{\Gamma \left( {{m_x}} \right)\Gamma \left( {{m_y}} \right)\Gamma \left( {{m_z}} \right)\Gamma \left( {{m_w}} \right)\Gamma \left( {{m_z}} \right){{\left( {\frac{{{{\bar \gamma }_{\rm{R}}}{{\bar \gamma }_{\rm{P}}}}}{{{{\bar \gamma }_{\rm{S}}}}}{{\bar \gamma }_y}} \right)}^{{m_Y}}}{{\left( {{{\bar \gamma }_{\rm{P}}}{{\bar \gamma }_z}} \right)}^{2{m_z}}}}} \times } \nonumber\\&
\Scale[0.98]{\sum\limits_{n = 0}^{{m_x} - 1} {\sum\limits_{{i_1} = 0}^n {\sum\limits_{{i_2} = 0}^{n - {i_1}} {\sum\limits_{k = 0}^{{m_w} - 1} {\sum\limits_{{k_1} = 0}^k {\left( \begin{array}{l}
k\\
{k_1}
\end{array} \right)} } \left( \begin{array}{l}
n\\
{i_1}
\end{array} \right)} } \left( \begin{array}{l}
n - {i_1}\\
\,\,\,{i_2}
\end{array} \right)\frac{{{{\left( {\frac{{{{\bar \gamma }_{\rm{R}}}}}{{{{\bar \gamma }_{\rm{S}}}}} + 1} \right)}^{n - {i_1} - {i_2}}}{{\left( {\frac{{{m_x}}}{{{{\bar \gamma }_x}{{\bar \gamma }_{\rm{R}}}}}} \right)}^n}{{\left( {\frac{{{m_w}}}{{{{\bar \gamma }_W}{{\bar \gamma }_{\rm{R}}}}}} \right)}^k}}}{{n!k!}}}  \times } \nonumber\\&
\Scale[0.98]{\left( {{m_y} + {i_1} - 1} \right)!\left( {{m_z} + {i_2} - 1} \right)!\left( {{m_z} + {k_1} - 1} \right)!\, \times {\left( {\frac{{{m_w}}}{{{{\bar \gamma }_W}{{\bar \gamma }_{\rm{R}}}}}} \right)^{ - \left( {{m_z} + {k_1}} \right)}}{\left( {\frac{{{m_x}}}{{{{\bar \gamma }_x}{{\bar \gamma }_{\rm{R}}}}}} \right)^{ - \left( {{m_z} + {i_2}} \right)}}{\left( {\frac{{{m_x}}}{{{{\bar \gamma }_x}{{\bar \gamma }_{\rm{R}}}}}} \right)^{ - \left( {{m_y} + {i_1}} \right)}}} \times  \nonumber\\&
\Scale[0.98]{ \times \left[ {\sum\limits_{{j_1} = 1}^{{m_z} + {k_1}} {{A_{{j_1}}}} \Gamma \left( {n + k + \frac{1}{2}} \right){\alpha _1}^{n + k + \frac{1}{2} - {j_1}}\Psi \left( {n + k + \frac{1}{2},n + k - \frac{1}{2} - {j_1},\left( {\frac{{\left( {\frac{{{{\bar \gamma }_{\rm{R}}}}}{{{{\bar \gamma }_{\rm{S}}}}} + 1} \right){m_x}}}{{{{\bar \gamma }_x}{{\bar \gamma }_{\rm{R}}}}} + \frac{{{m_w}}}{{{{\bar \gamma }_W}{{\bar \gamma }_{\rm{R}}}}} + b} \right){\alpha _1}} \right)} \right.} \nonumber\\&
\Scale[0.98]{ + \sum\limits_{{j_2} = 1}^{{m_z} + {i_2}} {{A_{{j_2}}}} \Gamma \left( {n + k + \frac{1}{2}} \right){\alpha _1}^{n + k + \frac{1}{2} - {j_2}}\Psi \left( {n + k + \frac{1}{2},n + k - \frac{1}{2} - {j_2},\left( {\frac{{\left( {\frac{{{{\bar \gamma }_{\rm{R}}}}}{{{{\bar \gamma }_{\rm{S}}}}} + 1} \right){m_x}}}{{{{\bar \gamma }_x}{{\bar \gamma }_{\rm{R}}}}} + \frac{{{m_w}}}{{{{\bar \gamma }_W}{{\bar \gamma }_{\rm{R}}}}} + b} \right){\alpha _2}} \right)} \nonumber\\&
\Scale[0.98]{\left. { + \sum\limits_{{j_3} = 1}^{{m_y} + {i_1}} {{A_{{j_3}}}} \Gamma \left( {n + k + \frac{1}{2}} \right){\alpha _1}^{n + k + \frac{1}{2} - {j_3}}\Psi \left( {n + k + \frac{1}{2},n + k - \frac{1}{2} - {j_3},\left( {\frac{{\left( {\frac{{{{\bar \gamma }_{\rm{R}}}}}{{{{\bar \gamma }_{\rm{S}}}}} + 1} \right){m_x}}}{{{{\bar \gamma }_x}{{\bar \gamma }_{\rm{R}}}}} + \frac{{{m_w}}}{{{{\bar \gamma }_W}{{\bar \gamma }_{\rm{R}}}}} + b} \right){\alpha _3}} \right)} \right]},
 \label{SEP_Final}
‎\end{align}‎
\rule{\linewidth}{1pt}
\vspace{-3mm}
\end{figure*}
\end{theorem}
where ($a$,$b$) are constants specified by the type of modulation; $\Gamma \left( . \right)$‎ is the gamma function~\cite[Eq. (8.310.1)]{Gradshteyn:Table:2007}, $\Psi \left( {.,.,.} \right)$ is the Tricomi confluent hypergeometric function~\cite[Eq. (9.210.2)]{Gradshteyn:Table:2007}; ${A_{{m_z+k_1-j_1+1}}} ‎\hspace{-1mm}‎=\hspace{-1mm} \frac{{\psi _1^{\left( {{m_z} + {k_2} - {j_1}} \right)}\left( { - {\alpha _1}} \right)}}{{\left( {{j_1} - 1} \right)!}}$, ${A_{{j_2}}} \hspace{-1mm}=\hspace{-1mm} \frac{{\psi _2^{\left( {{m_z} + {i_2} - {j_2}} \right)}\left( { - {\alpha _2}} \right)}}{{\left( {{m_z} + {i_2} - {j_2}} \right)!}}$, ${A_{{j_3}}} \hspace{-1mm}=\hspace{-1mm} \frac{{\psi _3^{\left( {{m_y} + {i_1} - {j_3}} \right)}\left( { - {\alpha _3}} \right)}}{{\left( {{m_y} + {i_1} - {j_3}} \right)!}}$; ${\alpha _1} \hspace{-1mm}=\hspace{-1mm} \frac{{{{\bar \gamma }_W}{{\bar \gamma }_{\rm{R}}}{m_z}}}{{{{\bar \gamma }_{\rm{P}}}{{\bar \gamma }_z}{m_w}}}$, ${\alpha _2} \hspace{-1mm}=\hspace{-1mm} \frac{{{{\bar \gamma }_x}{{\bar \gamma }_{\rm{R}}}{m_z}}}{{{{\bar \gamma }_{\rm{P}}}{{\bar \gamma }_z}{m_x}}}$, $\,{\alpha _3} \hspace{-1mm}=\hspace{-1mm} \frac{{{{\bar \gamma }_x}{{\bar \gamma }_{\rm{S}}}{m_y}}}{{{{\bar \gamma }_{\rm{P}}}{{\bar \gamma }_y}{m_x}}}$, ${\psi _1}\left( \Theta  \right) = \frac{1}{{{{\left( {\Theta  + {\alpha _2}} \right)}^{{m_z} + {i_2}}}{{\left( {\Theta  + {\alpha _3}} \right)}^{{m_y} + {i_1}}}}}$, ${\psi _2}\left( \Theta  \right) \hspace{-1mm}=\hspace{-1mm} \frac{1}{{{{\left( {\Theta  + {\alpha _1}} \right)}^{{m_z} + {k_1}}}{{\left( {\Theta  + {\alpha _3}} \right)}^{{m_y} + {i_1}}}}}$ and ${\psi _3}\left( \Theta  \right) \hspace{-1mm}=\hspace{-1mm} \frac{1}{{{{\left( {\Theta  + {\alpha _1}} \right)}^{{m_z} + {k_1}}}{{\left( {\Theta  + {\alpha _2}} \right)}^{{m_z} + {i_2}}}}}$.
\proof
\emph{In order to facilitate error derivation‎, \eqref{CDF} should be addressed in a more mathematically tractable form‎. ‎For this purpose‎, by substituting \eqref{CDF} into ~\cite[Eq. (9)]{Duong:Let:2013} and ‎utilizing partial fraction expansions~\cite[Eq. (2.102)]{Gradshteyn:Table:2007}‎, and based on~\cite[Eq. (9.210.2)]{Gradshteyn:Table:2007},~\cite[Eq. (3.361.2)]{Gradshteyn:Table:2007}, a closed-form expression for the SEP over i.n.i.d. Nakagami-$m$ fading channels can be derived as‎ in \eqref{SEP_Final}.}

\vspace{-6mm}
‎\section{Simulation Results and Discussions}‎
‎In this section‎, ‎we perform Monte carlo simulation to verify the accuracy of our analysis where 4PSK is employed for the data symbols and assuming ${\gamma _{{\rm{th}}}} = 3\,{\rm{dB}}$ to determine the OC. Monte-Carlo simulations are averaged over $10^5$ independent one.

In Fig‎. ‎1, we present the effect of primary transmit SNR on the
secondary OC of Scenario (a) for fading severity parameter equal to $m = 2$. ‎In the same figure, the simulation results of the OC is depicted along with the results related to the lower bound‎, ‎given in \eqref{OP-Scenario a} for different values of ${P_{{\rm{ou}}{{\rm{t}}_{{\rm{Pri}}}}}^{\rm{Thr}}}$. ‎It can be noticed that the decrease in threshold (${P_{{\rm{ou}}{{\rm{t}}_{{\rm{Pri}}}}}^{\rm{Thr}}}$) enhance the outage performance of the PUs, in turn, provides an intensive limitation for transmit powers of SUs which further leads in an increase of the secondary OC. However, if SNR of the primary transmitter is increased beyond a certain level, the maximum allowed power is reached for SUs, which does not allow further increase in $P_{S_i}$ and $P_R$. Thus, with an additional increase in the primary power, an error floor occurs for the outage performance of the secondary network.

In Fig‎. ‎2‎, ‎we analyze the SEP of Scenario (a)‎. It is worth noting that an increase in ${P_{{\rm{ou}}{{\rm{t}}_{{\rm{Pri}}}}}^{\rm{Thr}}}$ relaxes the limitation of the power of the SUs. However, if we increase ${P_{{\rm{ou}}{{\rm{t}}_{{\rm{Pri}}}}}^{\rm{Thr}}}$ beyond a certain level, the power of the SU nodes will reach it maximum admissible power. Hence, in such a case regardless of the value of the ${P_{{\rm{ou}}{{\rm{t}}_{{\rm{Pri}}}}}^{\rm{Thr}}}$ the secondary outage performance remains constant. We can see that our analytical results match well with the simulation results, which validates our theoretical analysis. In addition, the outage performance improves with the increase of the fading severity parameters (compared with \cite{Vahidian1:TVT:2016} where $m=1$).

In Fig‎.~3, we present the outage performance of Scenario (b) versus the SNR‎. ‎In this figure the simulation results of the error probability is illustrated along with the results pertaining to the lower bound‎, ‎given in \eqref{OP-Scenario b} for various values of ${P_{{\rm{ou}}{{\rm{t}}_{{\rm{Pri}}}}}^{\rm{Thr}}}$. As can be seen the outage performance improves when the number of relays increases specially, in the median and high SNR regions. ‎As observed‎, ‎the error and outage performance of cognitive radio network in Scenario (a) and (b), deteriorate with the decrease of ${P_{{\rm{ou}}{{\rm{t}}_{{\rm{Pri}}}}}^{\rm{Thr}}}$. More importantly, when ‎strict limits on the transmit power of the secondary nodes is imposed (${P_{{\rm{ou}}{{\rm{t}}_{{\rm{Pri}}}}}^{\rm{Thr}}}=0$), in such a case, no transmission allowed for the secondary network, therefore, the OC of the secondary network equals one. Furthermore, in Scenario (b) the selection diversity enables the secondary system to perform better by changing both the choice of the relay and its transmit power depending on the channel fades.

\vspace{-6mm}
‎\section{Conclusion}‎
In this paper, we have developed new analytical results that can be used to investigate the OC and ASEP of non-regenerative cognitive two-way relaying networks with underlay spectrum sharing in Nakagami-$m$ fading environment. Two scenarios are treated including Scenario (a) in which two transmit nodes communicate via a single relay, or communicate through a selected relay as considered in Scenario (b). For both scenarios, tight closed-form expressions for OC were derived which provided an efficient means to investigate the impact of the PU's interference and fading parameters on the outage performance of the system. Moreover, a lower bound expression of ASEP for Scenario (a), were derived. Simulations results were provided to testify the correctness of the derive analytical results and to study the effect of many parameters such as the number of relays, constellation size, and channel statistics on the considered performance measures.



\vspace{4.5mm}
\begin{figure}[h]
\centering
\includegraphics[width=70mm, height=40mm]{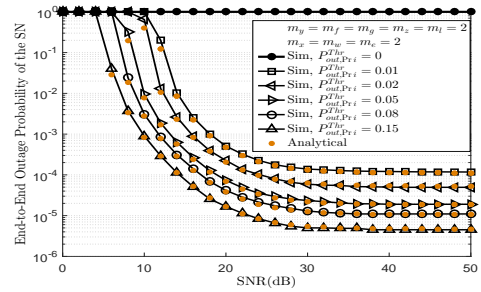}
\vspace{-1 em}
\caption{OC of the cognitive two-way non-regenerative relaying for different values of ${P_{{\rm{ou}}{{\rm{t}}_{{\rm{Pri}}}}}^{\rm{Thr}}}$ and for $m=2$.}
\end{figure}
\vspace{-4mm}
\begin{figure}[h]
\centering
\hspace{-12mm}\includegraphics[width=75mm, height=40mm]{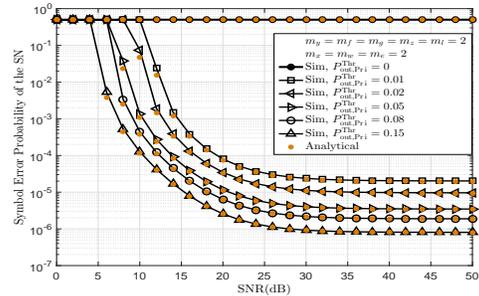}
\vspace{-1 em}
\caption{SEP of the two-way cognitive radio network for various values of primary OC threshold ${P_{{\rm{ou}}{{\rm{t}}_{{\rm{Pri}}}}}^{\rm{Thr}}}$ and for $m=2$.}
\end{figure}

\vspace{-4mm}
\begin{figure}[h]
\centering
\hspace{-12mm}\includegraphics[width=80mm, height=40mm]{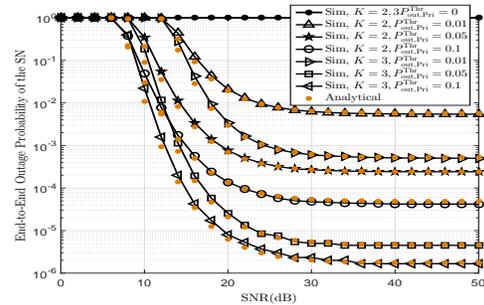}
\vspace{-1 em}
\caption{End-to-end OC of the cognitive radio network with best relay selection strategy for different number of relays and different values of ${P_{{\rm{ou}}{{\rm{t}}_{{\rm{Pri}}}}}^{\rm{Thr}}}$ and for $m=2$.}
\vspace{-1.2 em}
\end{figure}

\appendices
\vspace{-.2mm}
\section{Proof of Theorem~\ref{Thm-outage probability-DMM}}
\label{apx1}
We assume the following change of variables
${\left| {{h_{R - {S_2}}}} \right|^2} = X$, ${\left| {{h_{R - {S_1}}}} \right|^2} = W$, $\frac{{{{\bar \gamma }_{\rm{R}}}{{\bar \gamma }_{\rm{P}}}}}{{{{\bar \gamma }_{\rm{S}}}}}{\left| {{h_{PT - R}}} \right|^2} = Y$, ${\bar \gamma _{\rm{P}}}{\left| {{h_{PT - {S_1}}}} \right|^2} = Z$, ${\bar \gamma _{\rm{P}}}{\left| {{h_{PT - {S_2}}}} \right|^2} = V,{\left| {{h_{PT - PX}}} \right|^2} = E$, ${\left| {{h_{{S_1} - PX}}} \right|^2} = F$, ${\left| {{h_{{S_2} - PX}}} \right|^2} = G$ and ${\left| {{h_{R - PX}}} \right|^2} = L$. The OC of the primary network can be obtained as

 ‎\begin{align}‎
&{P_{out,\Pr i}} \hspace{-1mm}=\hspace{-1mm} \left\{ {\log \left( {1 \hspace{-1mm}+\hspace{-1mm} \frac{{{{\bar \gamma }_P}{{\left| {{h_{PT \hspace{-0.5mm}-\hspace{-0.5mm} PX}}} \right|}^2}}}{{{{\bar \gamma }_{{S_1}}}{{\left| {{h_{{S_1} \hspace{-0.5mm}-\hspace{-0.5mm} PX}}} \right|}^2} \hspace{-1mm}+\hspace{-0.9mm} {{\bar \gamma }_{{S_2}}}{{\left| {{h_{{S_2} \hspace{-0.5mm}- \hspace{-0.5mm} PX}}} \right|}^2} \hspace{-1.1mm}+\hspace{-1.2mm} 1}}} \right)\hspace{-1mm} \le\hspace{-1.2mm} {R_P}} \right\}\nonumber
\end{align}
\vspace{-3mm}
 ‎\begin{align}‎
& = {{\rm{E}}_{F,G}}\left[ {\Pr \left\{ {E \le \frac{\Theta }{{{{\bar \gamma }_{\rm{P}}}}}\left( {{{\bar \gamma }_{{{\rm{S}}_1}}}F + {{\bar \gamma }_{{{\rm{S}}_2}}}G + 1} \right)\left| {F,G} \right.} \right\}} \right] = \nonumber\\&
{{\rm{E}}_{F,G}}\left[ {{F_E}\left( {\frac{\Theta }{{{{\bar \gamma }_{\rm{P}}}}}\left( {{{\bar \gamma }_{{{\rm{S}}_1}}}F + {{\bar \gamma }_{{{\rm{S}}_2}}}G + 1} \right)\left| {F,G} \right.} \right)} \right].
\end{align}

The cdf of a Gamma RV, $X$, is defined as

 ‎\begin{align}‎
F_X\left( x \right)=\frac{{\Gamma \left( {{m_x},\left( {\frac{{{m_x}}}{{{{\bar \gamma }_x}}}} \right)} \right)}}{{\Gamma \left( {{m_x}} \right)}}\label{cd}.
\end{align}
Therefore, we have
 ‎\begin{align}‎
\Scale[0.97]{{P_{out,\Pr i}} \hspace{-1mm}=\int {\int_{Z,Y>0} {} } 1 - }&\Scale[0.97]{\frac{{\Gamma \left( {{m_e},\left( {\frac{{{m_e}}}{{{{\bar \gamma }_e}}}} \right)\frac{\Theta }{{{{\bar \gamma }_{\rm{P}}}}}\left( {{{\bar \gamma }_{{{\rm{S}}_1}}}F + {{\bar \gamma }_{{{\rm{S}}_2}}}G + 1} \right)} \right)}}{{\Gamma \left( {{m_e}} \right)}}}\nonumber\\& \times {f_F}\left( f \right){f_G}\left( g \right) dFdG,\label{pdf-3}
\end{align}
 with the pdf of the RVs $F$ and $G$ at hand as in
  ‎\begin{align}‎
&\Scale[0.97]{{f_F}\left( f \right) = \frac{{{{\left( {{m_f}} \right)}^{{m_f}}}{F^{{m_f} - 1}}}}{{\Gamma \left( {{m_f}} \right)\bar \gamma _f^{{m_f}}}}\exp \left( {\frac{{ - F{m_f}}}{{{{\bar \gamma }_f}}}} \right)},\notag\\& \Scale[0.97]{{f_G}\left( g \right) = \frac{{{{\left( {{m_g}} \right)}^{{m_g}}}{G^{{m_g} - 1}}}}{{\Gamma \left( {{m_g}} \right){{\left( {{{\bar \gamma }_g}} \right)}^{{m_g}}}}}\exp \left( {{\raise0.7ex\hbox{${ - G{m_g}}$} \!\mathord{\left/
 {\vphantom {{ - G{m_g}} {{{\bar \gamma }_{\rm{P}}}{{\bar \gamma }_g}}}}\right.\kern-\nulldelimiterspace}
\!\lower0.7ex\hbox{${{{\bar \gamma }_{\rm{P}}}{{\bar \gamma }_g}}$}}} \right)},\label{pdf-2}
\end{align}

\noindent and by substituting \eqref{pdf-2} in \eqref{pdf-3} and making use of the identities~\cite{Gradshteyn:Table:2007} as in the following we arrive at \eqref{OP of PN}.

\begin{align}‎
{\left( {{\varepsilon _1} + {\varepsilon _2} + {\varepsilon _3}} \right)^n} = \sum\limits_{{i_1} = 0}^n {\sum\limits_{{i_2} = 0}^{n - {i_1}} {\left( \begin{array}{l}
n\\
{i_1}
\end{array} \right)} } \left( \begin{array}{l}
n - {i_1}\\
\,\,\,{i_2}
\end{array} \right){\varepsilon _1^{{i_2}}}{\varepsilon _2^{{i_1}}}{\varepsilon _3^{n - {i_1} - {i_2}}},
\label{I1}
 \end{align}
 \vspace{-3mm}
 ‎\begin{align}‎
\int_0^\infty  {{x^s}\exp \left( { - hx} \right)dx = s!{{\left( {\frac{1}{h}} \right)}^{s + 1}}}  = s!{\left( h \right)^{ - \left( {s + 1} \right)}},\label{I2}
 \end{align}
\vspace{-3mm}
 ‎\begin{align}‎
\Gamma \left( {1 + n,x} \right) = n!{e^{ - x}}\sum\limits_{m = 0}^n {\frac{{{x^m}}}{{m!}}}. \label{I3}
\end{align}

\vspace{-5mm}

\section{Proof of Theorem~\ref{op-secanrio-a}}
\label{apx2}
\vspace{-5mm}
 ‎\begin{align}‎
&\Scale[0.97]{{F_{\gamma _{{{S1}}}^{{\rm{up}}}}}\left( \Theta  \right) = \Pr \left\{ {{{\bar \gamma }_{\rm{R}}}\min \left( {\frac{X}{{Z + Y + \frac{{{{\bar \gamma }_{\rm{R}}}}}{{{{\bar \gamma }_{\rm{S}}}}} + 1}},\frac{{\rm{W}}}{{Z + 1}}} \right) < \Theta } \right\} \approx  1 -}\nonumber\\&
 \Scale[0.94]{\underbrace {{{\rm{E}}_{Z,Y}}\hspace{-2mm}\left[\hspace{-1mm} {\Pr \left\{ {\frac{X}{{Z \hspace{-1mm}+\hspace{-1mm} Y \hspace{-1mm}+\hspace{-1mm} \frac{{{{\bar \gamma }_{\rm{R}}}}}{{{{\bar \gamma }_{\rm{S}}}}} +\hspace{-1mm} 1}} >\hspace{-1mm} \frac{\Theta }{{{{\bar \gamma }_{\rm{R}}}}}\left| Z \right.,Y} \right\}} \right]}_{{\chi _1}} \hspace{-1mm}\times \hspace{-1mm}\underbrace {{{\rm{E}}_Z}\left[ {\Pr \left\{ {\frac{{\,{\rm{W}}}}{{Z \hspace{-1mm}+\hspace{-1mm} 1}} > \hspace{-1mm}\frac{\Theta }{{{{\bar \gamma }_{\rm{R}}}}}\left| Z \right.} \right\}} \right]}_{{\chi _2}}}.\label{we}
\end{align}

First, we focus on evaluating ${{\chi _1}}$ which with the concepts of probability and with the help of \eqref{cd} as well as the pdf of the RVs $Z$ and $Y$ can be given as
 ‎\begin{align}‎
{\chi _1}\hspace{-1mm} =&\hspace{-1mm} \int \hspace{-1mm}{\int_{Z,Y>0} {} } \hspace{-4mm}  \frac{{\Gamma \left( {{m_x},\left( {\frac{{{m_x}}}{{{{\bar \gamma }_x}}}} \right)\left( {Z \hspace{-1mm}+ \hspace{-1mm}Y \hspace{-1mm}+\hspace{-1mm} \frac{{{{\bar \gamma }_{\rm{R}}}}}{{{{\bar \gamma }_{\rm{S}}}}} \hspace{-1mm}+ \hspace{-1mm}1} \right)\frac{\Theta }{{{{\bar \gamma }_{\rm{R}}}}}} \right)}}{{\Gamma \left( {{m_x}} \right)}}\frac{{{{\left( {{m_Y}} \right)}^{{m_Y}}}{Y^{{m_Y} - 1}}}}{{\Gamma \left( {{m_Y}} \right){{\left( {\frac{{{{\bar \gamma }_{\rm{R}}}{{\bar \gamma }_{\rm{P}}}}}{{{{\bar \gamma }_{\rm{S}}}}}{{\bar \gamma }_y}} \right)}^{{m_Y}}}}}\nonumber\\&
\exp \left( {\frac{{ - Y{m_Y}}}{{\frac{{{{\bar \gamma }_{\rm{R}}}{{\bar \gamma }_{\rm{P}}}}}{{{{\bar \gamma }_{\rm{S}}}}}{{\bar \gamma }_Y}}}} \right)
\frac{{{{\left( {{m_z}} \right)}^{{m_z}}}{Z^{{m_z} - 1}}}}{{\Gamma \left( {{m_z}} \right){{\left( {{{\bar \gamma }_{\rm{P}}}{{\bar \gamma }_z}} \right)}^{{m_z}}}}}\exp \left( {\frac{{ - Z{m_z}}}{{{{\bar \gamma }_{\rm{P}}}{{\bar \gamma }_z}}}} \right)dYdZ.
\end{align}

Next, utilizing the identities in \eqref{I1}, and \eqref{I2}  and also after some calculations we achieve ${\chi _1}$. In addition, ${\chi _2}$ can be computed taking the same steps as that of ${\chi _1}$. Finally, by substituting the former and the latter in \eqref{we} we arrive at ${F_{\gamma _{{{S1}}}^{{\rm{up}}}}}\left( \Theta  \right)$.
\vspace{-4mm}
\section{Proof of Theorem~\ref{op-secanrio-b}}
\label{apx3}
\vspace{-7mm}
‎\begin{align}‎
&\Scale[0.97]{{F_{\gamma _{{\rm{e2e}}}^{{\rm{up}}}}}\left( \Theta  \right) =
\mathop {\max }\limits_{k = 1,2,...,K} \left\{ {\min \left( {\gamma _{{\rm{S_1}}}^{{\rm{up}}},\gamma _{{\rm{S_2}}}^{{\rm{up}}}} \right)} \right\} = }\nonumber\\&
\Scale[0.97]{\Pr \hspace{-1mm}\left[\hspace{-1mm} {\mathop {\max }\limits_{k = 1,2,...,K} \hspace{-1mm}\left\{ {\min \hspace{-1mm}\left\{ \hspace{-1mm}{\frac{{{{\bar \gamma }_{\rm{R}}}}}{{{{\bar \gamma }_{\rm{S}}}}}\min\hspace{-1mm} \left( \hspace{-1mm}{\frac{{{{\bar \gamma }_{\rm{S}}}{{\left| {{h_{{S_2} \hspace{-1mm}-\hspace{-1mm} {R_k}}}} \right|}^{\rm{2}}}}}{{\frac{{{{\bar \gamma }_{\rm{R}}}{{\bar \gamma }_{\rm{P}}}}}{{{{\bar \gamma }_{\rm{S}}}}}{h_{PT - {R_k}}} \hspace{-1mm}+\hspace{-1mm} \frac{{{{\bar \gamma }_{\rm{R}}}}}{{{{\bar \gamma }_{\rm{S}}}}} \hspace{-1mm}+\hspace{-1mm} 1}},\,{{\bar \gamma }_{{{\rm{S}}_1}}}{{\left| {{h_{{\rm{ST1 - S}}{{\rm{R}}_k}}}} \right|}^{\rm{2}}}} \right),}\hspace{-1mm} \right.} \hspace{-1mm}\right.}\hspace{-1mm} \right.}\nonumber\\&
\Scale[0.97]{\left. \hspace{-1mm}{\left. {\left. {\frac{{{{\bar \gamma }_{\rm{R}}}}}{{{{\bar \gamma }_{\rm{S}}}}}\min \left( {\frac{{{{\bar \gamma }_{\rm{S}}}{{\left| {{h_{{S_1} - {R_k}}}} \right|}^{\rm{2}}}}}{{\frac{{{{\bar \gamma }_{\rm{R}}}{{\bar \gamma }_{\rm{P}}}}}{{{{\bar \gamma }_{\rm{S}}}}}{h_{PT - {R_k}}} + \frac{{{{\bar \gamma }_{\rm{R}}}}}{{{{\bar \gamma }_{\rm{S}}}}} + 1}},\,{{\bar \gamma }_{\rm{S}}}{{\left| {{h_{{{\rm{S}}_{\rm{2}}}{\rm{ - }}{{\rm{R}}_k}}}} \right|}^{\rm{2}}}} \right) \hspace{-1mm}<\hspace{-1mm} \Theta } \right\}} \right\}} \right].}
\end{align}

Let ${\cal{V}}=\min \left( {\gamma _{{\rm{S_1}}}^{{\rm{up}}},\gamma _{{\rm{S_2}}}^{{\rm{up}}}} \right)$. Due to the independency of the RVs, we first proceed to obtain the cdf of the RV ${\cal{V}}$ as in the following.
 ‎\begin{align}
&\Scale[0.97]{{F_{\cal{V}}}\left( \Theta \right)=1 - {E_{{Y_k}}}\Bigg[ \Pr \left( {{X_k} \ge \frac{{\left( {Y + \frac{{{{\bar \gamma }_{\rm{R}}}}}{{{{\bar \gamma }_{\rm{S}}}}} + 1} \right)\Theta }}{{{{\bar \gamma }_{\rm{R}}}}}\,\,,\,\,{X_k} \ge \frac{\Theta }{{{{\bar \gamma }_{{{\rm{R}}_k}}}}}} \right)}\nonumber\\&\Scale[0.97]{ \times
\Pr \left( {{W_k} \ge \frac{{\left( {Y_k + \frac{{{{\bar \gamma }_{\rm{R}}}}}{{{{\bar \gamma }_{\rm{S}}}}} + 1} \right)\Theta }}{{{{\bar \gamma }_{\rm{R}}}}}\,\,,\,\,{W_k} \ge \frac{\Theta }{{{{\bar \gamma }_{\rm{R}}}}}} \right) \Bigg]}\label{cdv}.
\end{align}
Using concepts of probability and having \eqref{cd} in mind, we can rewrite \eqref{cdv} as

\begin{align}
&\Scale[0.97]{{F_{\cal{V}}}\left( \Theta  \right) \hspace{-1mm}= \hspace{-1mm}1 \hspace{-1mm}-\hspace{-1mm} {E_{{Y_k}}}\left[ {{\Gamma ^{ - 1}}\left( {{m_{{x_k}}}} \right)\Gamma \left( {{m_{{x_k}}},\left( {\frac{{{m_{{x_k}}}}}{{{{\bar \gamma }_{{x_k}}}}}} \right)\frac{{\left( {{Y_k} + \frac{{{{\bar \gamma }_{\rm{R}}}}}{{{{\bar \gamma }_{\rm{S}}}}} + 1} \right)\Theta }}{{{{\bar \gamma }_{\rm{R}}}}}} \right)} \right.}\nonumber\\& \Scale[0.97]{\times
\left. {{\Gamma ^{ - 1}}\left( {{m_{{w_k}}}} \right)\Gamma \left( {{m_{{w_k}}},\left( {\frac{{{m_{{w_k}}}}}{{{{\bar \gamma }_{{w_k}}}}}} \right)\frac{{\left( {{Y_k} + \frac{{{{\bar \gamma }_{\rm{R}}}}}{{{{\bar \gamma }_{\rm{S}}}}} + 1} \right)\Theta }}{{{{\bar \gamma }_{\rm{R}}}}}} \right)} \right]}.
\end{align}
Making use of the identity in \eqref{I3}, we have
 ‎\begin{align}
&\Scale[0.98]{{F_{\cal{V}}}\left( \Theta \right)=\hspace{-1mm}
1 \hspace{-1mm}-\hspace{-1mm} {E_{Y_k}}\left[ {\left( {{m_{{x_k}}}\hspace{-1mm} -\hspace{-1mm} 1} \right)!{{\left( {\Gamma \left( {{m_{{x_k}}}} \right)} \right)}^{ - 1}}\left( {{m_{{w_k}}}\hspace{-1mm} - \hspace{-1mm}1} \right)!{{\left( {\Gamma \left( {{m_{{w_k}}}} \right)} \right)}^{ - 1}}} \right.}\nonumber\\&
\Scale[0.95]{\exp \left( { - \left( {\frac{{{m_{{x_k}}}}}{{{{\bar \gamma }_{{x_k}}}{{\bar \gamma }_{\rm{R}}}}}} \right)\left( {\frac{{{{\bar \gamma }_{{{\rm{R}}_k}}}}}{{{{\bar \gamma }_{\rm{S}}}}} + 1} \right)\Theta } \right)\exp \left( { - \left( {\frac{{{m_{{w_k}}}}}{{{{\bar \gamma }_{{w_k}}}{{\bar \gamma }_{\rm{R}}}}}} \right)\left( {\frac{{{{\bar \gamma }_{\rm{R}}}}}{{{{\bar \gamma }_{\rm{S}}}}} + 1} \right)\Theta } \right)}\nonumber\\&
\Scale[0.98]{\exp \left( { - \left( {\frac{{{m_{{x_k}}}\Theta }}{{{{\bar \gamma }_{{x_k}}}{{\bar \gamma }_{\rm{R}}}}} + \frac{{{m_w}\Theta }}{{{{\bar \gamma }_{{w_k}}}{{\bar \gamma }_{{{\rm{R}}_k}}}}}} \right)Y} \right)\sum\limits_{n = 0}^{{m_{{x_k}}} - 1} {\sum\limits_{{n_1} = 0}^{{m_{{x_k}}} - 1} {\frac{1}{{{n_1}!n!}}{{\left( {\frac{{{m_{{w_k}}}\Theta }}{{{{\bar \gamma }_{{w_k}}}{{\bar \gamma }_{\rm{R}}}}}} \right)}^{{n_1}}}} }} \nonumber\\&
\Scale[0.98]{\left. {{{\left( {\frac{{{x_k}\Theta }}{{{{\bar \gamma }_{{x_k}}}{{\bar \gamma }_{\rm{R}}}}}} \right)}^n}{{\left( {{Y_k} + \frac{{{{\bar \gamma }_{\rm{R}}}}}{{{{\bar \gamma }_{\rm{S}}}}} + 1} \right)}^{n + {n_1}}}} \right]}.
\end{align}
With pdf of $Y_k$ at hand as in the following and evoking \eqref{I1}, we reach \eqref{CDF_Selection}.
 ‎\begin{align}
{f_{{Y_k}}}\left( {{y_k}} \right) = \frac{{{{\left( {{m_{{y_k}}}} \right)}^{{m_{{y_k}}}}}Y_k^{{m_{{y_k}}} - 1}}}{{\Gamma \left( {{m_{{y_k}}}} \right){{\left( {\frac{{{{\bar \gamma }_{\rm{R}}}{{\bar \gamma }_{\rm{P}}}}}{{{{\bar \gamma }_{\rm{S}}}}}{{\bar \gamma }_{{y_k}}}} \right)}^{{m_{{y_k}}}}}}}\exp \left( {\frac{{ - {Y_k}{{\bar \gamma }_{\rm{S}}}{m_{{y_k}}}}}{{{{\bar \gamma }_{\rm{R}}}{{\bar \gamma }_{\rm{P}}}{{\bar \gamma }_{{y_k}}}}}} \right).
\end{align}


\section*{Acknowledgment}
This work was supported by NPRP from the Qatar National Research Fund (a member of Qatar Foundation) under grant no. 8-1545-2-657.

\small
\bibliographystyle{IEEEtran}
\bibliography{refrence}
\end{document}